\documentclass[10pt,twocolumn]{article} 

\usepackage{simpleConference}

\usepackage[utf8]{inputenc} 
\usepackage[T1]{fontenc}    
\usepackage{hyperref}       
\usepackage{url}            
\usepackage{booktabs}       
\usepackage{amsfonts}       
\usepackage{nicefrac}       
\usepackage{microtype}      
\usepackage{cleveref}       
\usepackage{lipsum}         
\usepackage{graphicx}
\usepackage{doi}
\usepackage{authblk}

\usepackage{multicol}
\usepackage{soul}
\usepackage{caption}
\usepackage{array}
\usepackage{subcaption}
\usepackage{booktabs}
\usepackage[official]{eurosym}

\begin{document}
\title{A vision transformer-based framework for knowledge transfer from multi-modal to mono-modal lymphoma subtyping models}
\author[1]{Bilel Guetarni}
\author[1]{Feryal Windal}
\author[2]{Halim Benhabiles}
\author[3]{Marianne Petit}
\author[3]{Romain Dubois}
\author[3]{Emmanuelle Leteurtre}
\author[4]{Dominique Collard}
\affil[1]{Junia, UMR 8520, University of Lille, CNRS, Centrale Lille, Univerity of Polytechnique Hauts-de-France, F-59000, Lille, France}
\affil[2]{IMT Nord Europe, Institut Mines-Télécom, Univ. Lille, Centre for Digital Systems, F-59000 Lille, France}
\affil[3]{Department of Pathology of the University Hospital of Lille, France}
\affil[4]{LIMMS/CNRS-IIS, IRL 2820, The University of Tokyo, and is the E-Health Chairholder of Yncrea Méditerranée, Yncrea Ouest, Junia , Toulon, Brest, Lille, France}

\maketitle

\begin{abstract}
Determining lymphoma subtypes is a crucial step for better patient treatment targeting to potentially increase their survival chances.
In this context, the existing gold standard diagnosis method, which relies on gene expression technology, is highly expensive and time-consuming, making it less accessibility.
Although alternative diagnosis methods based on IHC (immunohistochemistry) technologies exist (recommended by the WHO), they still suffer from similar limitations and are less accurate.
Whole Slide Image (WSI) analysis using deep learning models has shown promising potential for cancer diagnosis, that could offer cost-effective and faster alternatives to existing methods.
In this work, we propose a vision transformer-based framework for distinguishing DLBCL (Diffuse Large B-Cell Lymphoma) cancer subtypes from high-resolution WSIs.
To this end, we introduce a multi-modal architecture to train a classifier model from various WSI modalities.
We then leverage this model through a knowledge distillation process to efficiently guide the learning of a mono-modal classifier.
Our experimental study conducted on a lymphoma dataset of 157 patients shows the promising performance of our mono-modal classification model, outperforming six recent state-of-the-art methods.
In addition, the power-law curve, estimated on our experimental data, suggests that with more training data from a reasonable number of additional patients, our model could achieve competitive diagnosis accuracy with IHC technologies.
Furthermore, the efficiency of our framework is confirmed through an additional experimental study on an external breast cancer dataset (BCI dataset).
\end{abstract}

\section{Introduction}
\label{sec:introduction}

Diffuse Large B-Cell Lymphoma (DLBCL) is a cancer of the immune system cells with a significant worldwide impact.
More specifically, in 2020,  544,352 new cases have been diagnosed and 259,793 deaths have been recorded \cite{Sung2021GlobalCS}.
Additionally, projections indicate a 7\% increase in both new cases and deaths in Western Europe from 2020 to 2025 \cite{Kanas2021EpidemiologyOD}.
It has been shown that distinguishing between DLBCL subtypes, namely ABC (activated B-cell-like) and GCB (germinal-center B-cell–like), is crucial for treatment selection and the vital prognosis of the patient \cite{Mareschal2015AccurateCO}.
Indeed, clinical trials using new targeted treatments have shown promising results in terms of efficiency and survival chance improvement, notably for the patients with the most aggressive subtype (ABC) \cite{Nowakowski2015LenalidomideCW, Nowakowski2021ROBUSTAP}.

Currently, the gold standard method for DLBCL subtyping is a molecular biology method named RT-MLPA (\textit{reverse transcriptase multiplex ligation-dependent probe amplification} \cite{Mareschal2015AccurateCO}).
Nevertheless, its high cost (600\euro{}\footnote{\label{note1}estimation from the Lille University Hospital}) and longtime response (35 days\footnotemark[1]) have maintained the exploitation of other diagnostic methods, notably the microscopic examination of the patient tissue by a pathologist.

The tissue examination can be performed on two types of markers, namely IHC\footnote{Alternative diagnosis method recommended by the WHO when RT-MLPA is not available.} (immunohistochemistry) and HES (hematoxylin-eosin-safran) markers.
In the case of IHC, the expression levels of different proteins are quantified by the pathologist, which permits the determination of the DLBCL subtype based on the Hans algorithm \cite{Hans2004ConfirmationOT}.
Although the response time of this latter method is relatively fast (2 to 3 days\footnotemark[1]) and its cost (90\euro{}\footnotemark[1]) is reasonable compared to the gold standard, it has an error rate of 20\% (misclassified cases).
Additionally, it requires the involvement of the pathologist for the tissue examination which represents a fastidious task, notably in daily routines.
In the case of HES markers, the pathologists analyze the morphology and structural organization of the tissue cells.
However, due to the high similarity of these features between the two subtypes and the lack of objective criteria, pathologists struggle to reach an agreement on their decision.
Compared to the IHC, the HES analysis is less expensive (it does not require protein markers) but suffers from a higher error rate (30-35\%) with respect to the gold standard.

With the rapid increase of data in medical centers, there has been a growing interest in applying deep learning methods to medical image analysis \cite{Mishro2021ASO, Xing2018DeepLI, Atwany2022DeepLT}.
In the field of histopathology, the slide of a tissue sample can be digitized into high-resolution images named Whole Slide Images (WSI).
Deep learning methods have shown unprecedented performance on many automatic tasks on these images \cite{Srinidhi2019DeepNN}: detection of several cancers (e.g., breast, lung, colon, etc. \cite{Lu2020DataefficientAW, Wang2020WeaklySD, Zhang2022DTFDMILDF}), nuclei segmentation \cite{Graham2018HoverNetSS} or case/image similarity \cite{Hashimoto2021CasebasedSI, Hegde2019SimilarIS, Riasatian2021FineTuningAT}.
Although several works have addressed lymphoma subtyping \cite{Achi2018AutomatedDO, Miyoshi2020DeepLS, Steinbuss2021DeepLF, Gaudio2018ComputerdrivenQI, Hashimoto2022SubtypeCO, Li2020ADL}, to the best of our knowledge, there is no prior work on automatic differentiation of DLBCL molecular subtypes based on WSI analysis.
Among the main reasons that could explain this observation is the scarcity of publicly available annotated datasets (data protection and privacy).
In addition, the characterization of the two subtypes of DLBCL from WSI is a challenging task.
Indeed, the diversity of the cancer location in the patient’s body and the complex nature of the tissue lead to high intra-class variability for each subtype.
Besides, the extremely high resolution of WSI ($100,000 \times 100,000$ pixels on average) prevents the direct application of existing pretrained CNN architectures and involves highly demanding computational requirements.

In this work, we propose a deep learning-based framework for distinguishing between the ABC and GCB subtypes of DLBCL cancer from high-resolution WSI.
Our framework involves the development of a multi-modal deep architecture, serving as a teacher model for classifying the cancer subtypes using various WSI modalities, namely IHC-BCL6, IHC-CD10, IHC-MUM1 and HES.
Additionally, we leverage the concept of knowledge distillation to build a mono-modal classifier, referred to as the student model, which shows promising performance for differentiating between the subtypes using a single modality, namely HES.
It is worth mentioning that having a subtyping method based only on HES modality is of high interest for the health care system (pathologists, patients, laboratories).
Indeed, removing the three IHC modalities will permit to (i) lighten the data acquisition process in the laboratories, (ii) accelerate patient diagnosis, and (iii) avoid the costs related to the production of these modalities.

The main contributions of this work are the following:
\begin{itemize}
	\item We exploit the concept of knowledge distillation for efficiently driving the learning of a classifier following a \textit{4-to-1 WSI modality} training process. In this sense, the teacher model is used to support the student one to extract the most relevant features for DLBCL subtype characterization. In addition, we propose a generic design of the framework, enabling its exploitation for any classification task that requires knowledge transfer from a multi-modal classifier to a mono-modal one.
	\item We designed and developed a deep architecture tailored for handling multi-modal inputs (four modalities of WSI) in a one-shot pass. To deal with the extremely high resolution ($100,000 \times 100,000$ px) of these inputs, we incorporated in our architecture a ViT encoder that permits processing them in the form of a sequence of patches.
	\item We designed and developed a module offering a multi-modal features fusion mechanism to efficiently learn a discriminative representation of the two cancer subtypes.
	\item We conducted a comprehensive experimental study to demonstrate the efficiency of our method and its superiority compared to state-of-the-art approaches.
\end{itemize}

\section{Related works}
\label{sec:relatedworks}

Recently, Guidolin et al. \cite{Guidolin2021DifferentSD} have proposed an experimental study on 30 DLBCL patients, comprising 15 ABC and 15 GCB cases.
They aimed to quantify the spatial distribution of cells in WSIs using statistical measures, using five IHC protein markers for staining.
As raised by the authors, their experimental study has permitted to highlight that the combination of spatial statistics-derived parameters can be exploited for distinguishing between ABC and GCB.
However, the authors did not propose any automatic subtyping method.

Although the research on automated DLBCL subtyping is scarce, there is an extensive literature on utilizing deep neural networks for cancer classification from WSI \cite{Srinidhi2019DeepNN, Deng2020DeepLI, Campanella2019ClinicalgradeCP, Lu2020DataefficientAW, Zhang2022DTFDMILDF, Chen2022SelfSupervisedVT, Chen2022ScalingVT, Shao2021TransMILTB}.
Among these works, several ones have addressed the differentiation between DLBCL and other lymphoma types \cite{Achi2018AutomatedDO, Miyoshi2020DeepLS, Steinbuss2021DeepLF, Gaudio2018ComputerdrivenQI, Hashimoto2022SubtypeCO, Li2020ADL}.
For example, Hashimoto et al. \cite{Hashimoto2022SubtypeCO} proposed a CNN-based classification method to distinguish between three lymphoma cancers, namely DLBCL, AITL and CHL, from HES WSIs.
To deal with the high resolution of these images, they made the choice to subdivide them into a set of patches and adopted a Multiple Instance Learning (MIL) approach to train the final classifier.
To improve the performance of the classifier, they proposed a methodology that allows for the selection of the most relevant WSIs from the dataset during the training stage.
More specifically, they defined a typicality criterion that allows for the measurement of the relevance of each WSI in terms of cancer class representativeness.

In addition to the task of cancer classification from WSI, other works proposed a higher level of cancer analysis-related applications, namely patient survival prediction \cite{Nakhli2023AMIGOSM, Chen2022ScalingVT}.
In the case of \cite{Nakhli2023AMIGOSM}, the authors designed a multi-modal graph neural network architecture to train a model to predict survival chances in cases of bladder and ovarian cancer from WSI IHC modalities.
As raised by the authors, using a GNN enables the extraction of cell-centric features and processing of the WSI in a single forward pass.
In the case of Chen et al.\cite{Chen2022ScalingVT}, they proposed a backbone composed of a series of vision transformer units to encode multi-scale region features.
To this end, they choose to feed these units by exploiting the levels of a pyramidal representation of the WSI, where each level corresponds to a subdivision into patches of customized size.
To improve the performance of their model, they pretrained it by adopting a self-supervised learning strategy.

Other works have been proposed to address the diagnosis of different diseases from WSIs such as colorectal polyps classification \cite{Korbar2017DeepLF}, neurodegenerative disease diagnosis \cite{Kim2023DiagnosisOA} or virus infected cell identification \cite{Dong2017EvaluationsOD}.
For instance, the method proposed by Dwivedi et al. \cite{Dwivedi2022MultiSG} performs liver disease severity scoring from two WSIs modalities, namely Masson’s trichrome and HES.
For each modality, a CNN generates a heatmap of the WSI which is subsequently transformed into a graph representation.
Each heatmap is then processed by a graph neural network (GNN) to produce a vector that captures the characteristics of the WSI.
The disease severity score is predicted from the concatenation of the two modality vectors.

While previous studies \cite{Hu2020KnowledgeDF, Gao2019PrivilegedMD, Chen2021LearningWP, Wang2023PrototypeKD} have applied knowledge distillation for multi-modal to mono-modal knowledge transfer in the context of medical image segmentation, our work stands out as we have developed a unique approach tailored for tissue classification, a distinct task from segmentation, which handles high-resolution images.

\section{Methods}
\label{sec:methods}

\begin{figure}[h]
	\includegraphics[trim={8cm 1cm 4cm 1cm}, clip, width=\linewidth]{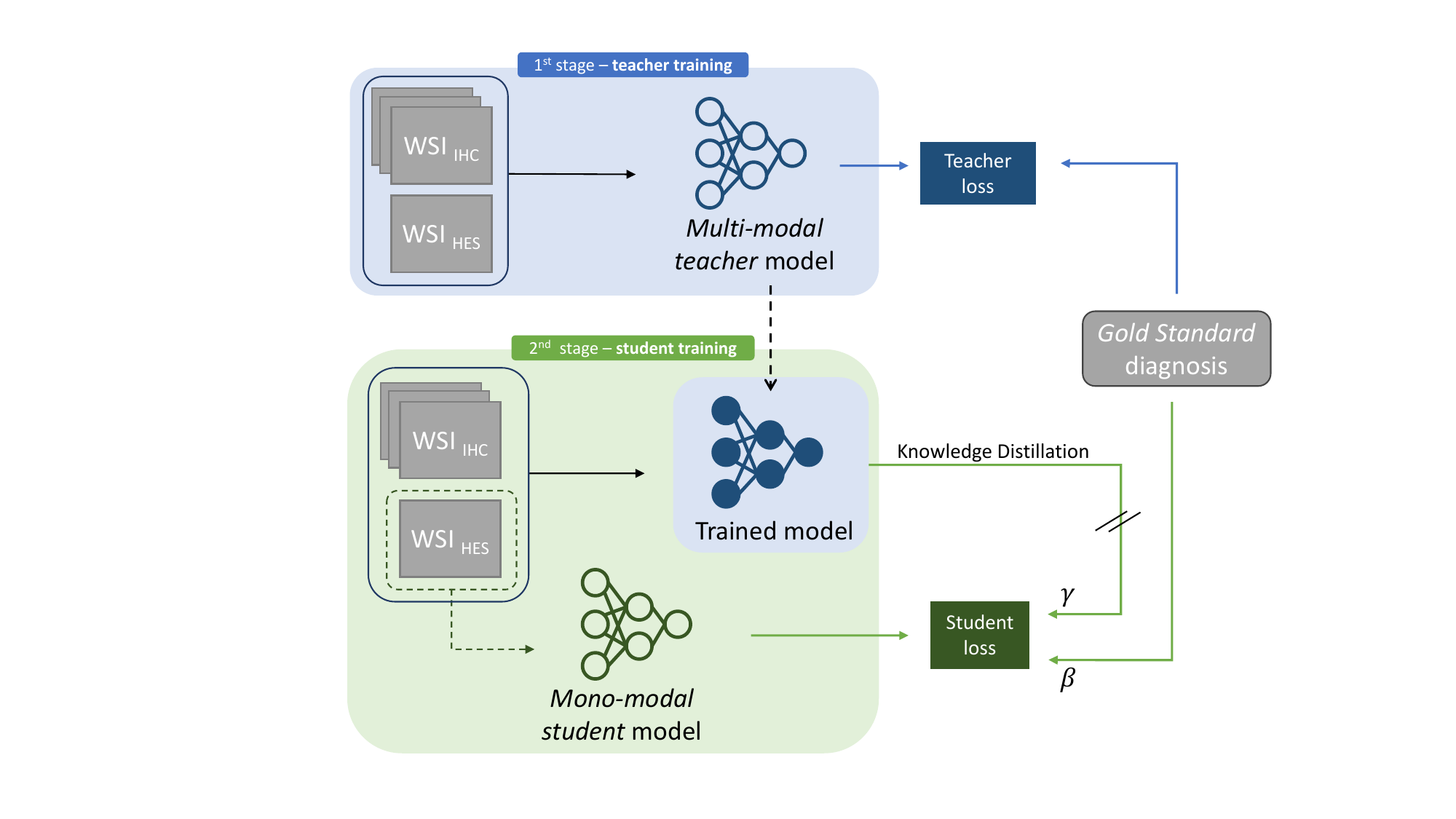}
	\caption{Knowledge Distillation for multi-to-mono modal WSI subtyping model.}
	\label{fig:methodology}
\end{figure}

\begin{figure*}[h]
	\includegraphics[trim={1cm 0cm 1cm 0cm}, clip, width=\textwidth]{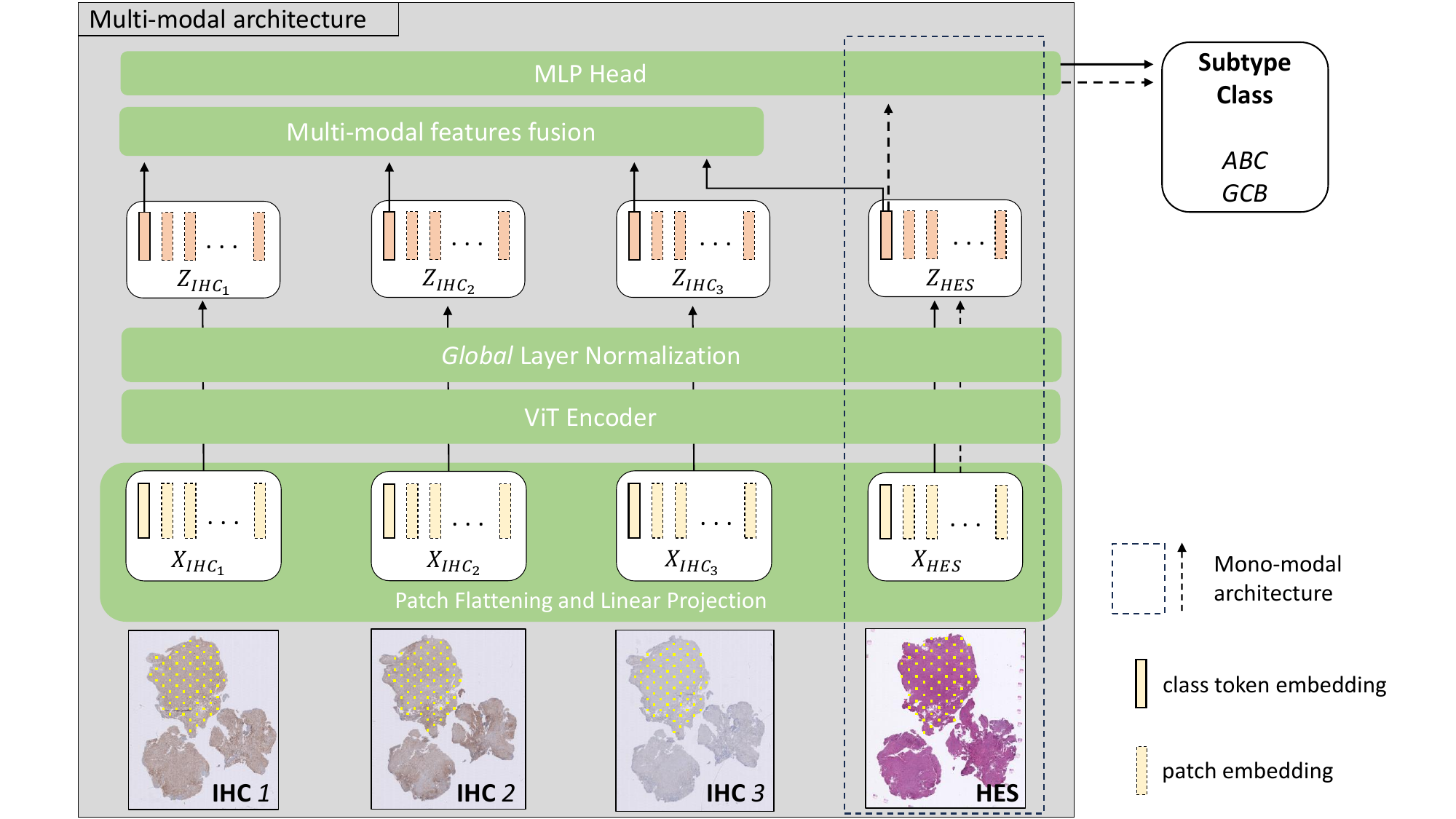}
	\caption{Our multi-modal architecture (teacher) takes as input a bag of sequences uniformly sampled from the same region across all WSI modalities (yellow squares). Its multi-modal features fusion mechanism produces a bag representation for multi-modal WSIs classification. Our mono-modal architecture is presented within the dotted bounding-box on the right.}
	\label{fig:architecture}
\end{figure*}

As illustrated in Fig. \ref{fig:methodology}, our framework designed for cancer subtyping from WSIs includes two classification models that are built by following a two-stage training process.
In the first stage, a multi-modal teacher classifier is trained on HES and IHC modalities.
In the second stage, the teacher acts through a knowledge distillation process as a robust supervision source for training a mono-modal student classifier on the HES modality.

Our deep learning architecture is summarized in Fig. \ref{fig:architecture}, and we next detail each component.

\subsection{From multi-slides input to single prediction}

In this section, we delve into the specifics of how our methodology processes WSIs stained with HES as well as IHC.

To avoid excessively complex computations related to the size of the regions, we adopted a region-based sampling approach.
It involves extracting matching regions of interest between all WSI modalities (see Fig. \ref{fig:architecture}).
For each region $R$ and each modality $m$, we extract from the region, RGB patches $x \in \mathbb{R}^{p \times p \times 3}$, with size $p \times p$, and we constructed sequences ($X$) of these patches as follows:

Let $N$ denote the number of patches extracted from R, and $S$ represents the number of patches in a sequence such that $X \in \mathbb{R}^{S \times p \times p \times 3}$.
We systematically extract sequences by randomly sampling $S$ patches uniformly distributed in R, without replacement.
This process is repeated until $N<S$ (i.e., there are less than $S$ patches left in the region).
Let $R_{m}$ be the set of all sequences extracted from region $R$ of modality $m$.
Let $B$ be a bag formed by sequences from all modalities, represented as:
\begin{equation} B := \{X_{m}, \forall m \in M\} \end{equation}
where $X_{m} \in R_{m}$ represents a sequence from modality $m$, and $M$ is the set of modalities.
Note that the sequences are sampled without replacement from each modality.
Indeed, when all sequences of one modality have been used, no more bags can be formed.
The bag $B$ serves as the input to the multi-modal model (i.e., teacher model).

Given that regions are represented as series of patches, we utilize the ViT architecture \cite{Dosovitskiy2020AnII} as our feature extractor.
Furthermore, this architecture has the capacity to learn inter-element relationships via its self-attention mechanism, which is of high interest in the case of WSI patches \cite{Chen2022ScalingVT, Chen2022SelfSupervisedVT}.
The ability to understand and leverage these relationships is crucial to learn global representation of WSIs.

To classify an image, the ViT architecture \cite{Dosovitskiy2020AnII} transforms the image into a sequence of patches.
These patches are subsequently flattened and projected by a linear layer into vectors known as patch embeddings.
In addition to this sequence, a learnable class token is prepended, serving as the image embedding post-encoder for the purpose of classification.
Our approach adheres to a similar structure: we flatten the patches of the sequences, project them with a linear layer, and prepend a class token.
However, in our scenario, we are working with a bag of sequences representing the same object but manifesting across different modalities (i.e., stains).
To account for variations in color distribution among these modalities, we employ a separate projection layer for each one.
The learnable class token embedding, however, is shared across all modalities.
The bag sequences are independently encoded by a shared ViT encoder and subsequently normalized with a Global Layer Normalization (GLN) \cite{Ba2016LayerN}.
All sequences $X_{m}$ of the bag $B$ are encoded and normalized simultaneously:
\begin{equation} Z_{m} = GLN(Encoder(X_{m})) \end{equation}
where $Encoder$ is the ViT encoder (including patch flattening and linear projection) and $GLN$ is the Global Layer Normalization.
We use the class token embeddings from every sequence after this GLN as input to our multi-modal features fusion mechanism, which produces the bag representation.
In our experimental study, we found that adding a softmax activation after the encoder's MLP improves the validation results.
In what follows, we refer to this modified version of the encoder unless otherwise stated.

\subsection{Multi-modal features fusion}

\begin{figure}[h]
	\includegraphics[trim={8cm 3cm 8cm 2cm}, clip, width=\linewidth]{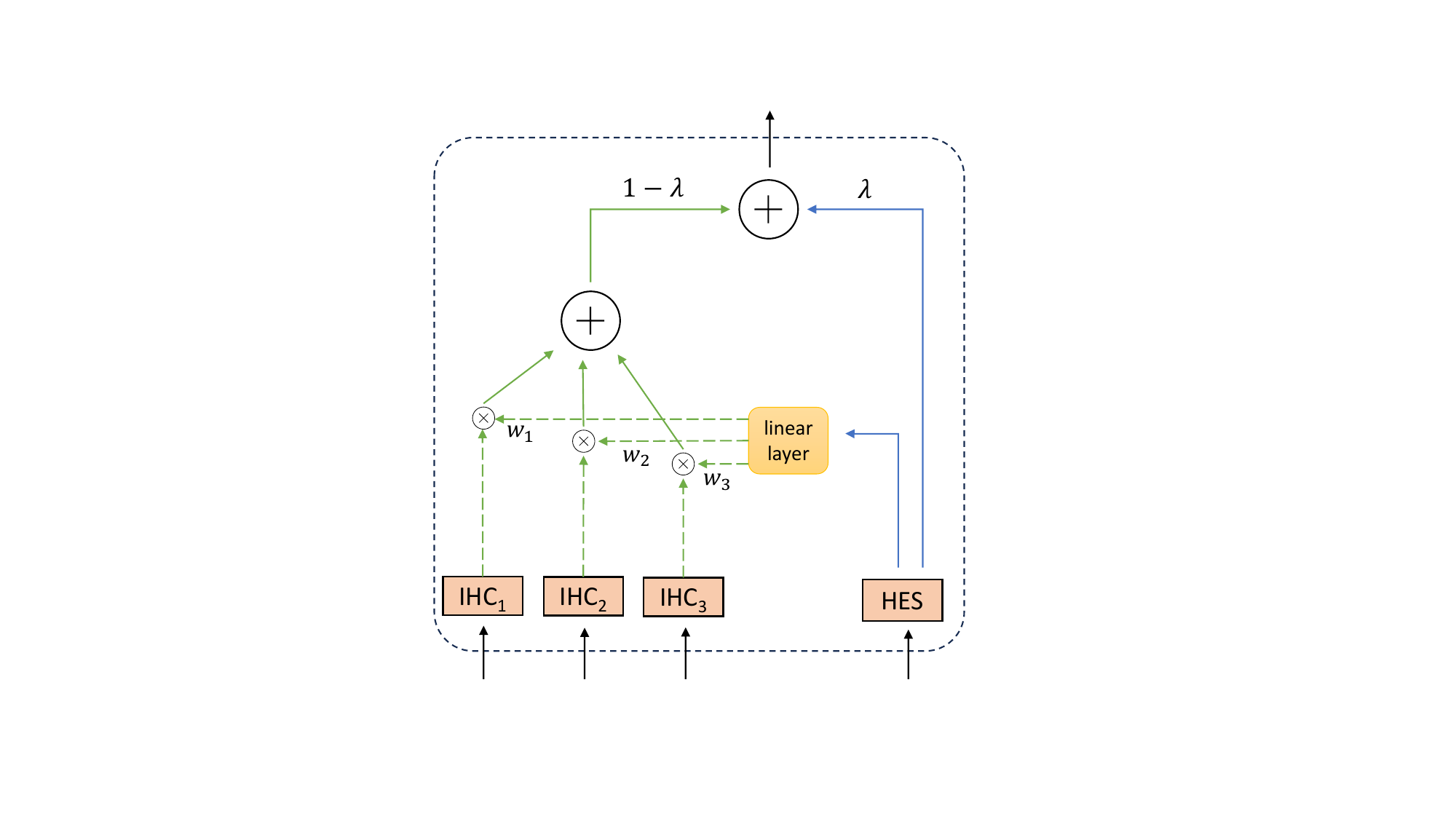}
	\caption{Our multi-modal features fusion mechanism.}
	\label{fig:crossstain}
\end{figure}

We design our multi-modal features fusion mechanism (Fig. \ref{fig:crossstain}) to fuse information from IHC and HES embeddings.
Because every modality is not equally important for different samples, we use a learnable attention weights approach to fuse IHC features before fusing it with HES.
For this, we consider HES as a query that is used to select information from IHC embeddings.
Whereas the current popular attention mechanism is defined as the dot-product attention between a query and keys \cite{Dosovitskiy2020AnII}, in our case, different modalities are involved, making the dot-product-based attention unsuitable. 
When the query and keys come from different space regions, it becomes less relevant to use the dot-product between the representation of the query and keys as a similarity indicator.
In our case, this is particularly evident as HES images exhibit a significantly different color distribution than IHC images (see inputs in Fig. \ref{fig:architecture}).
In this regard, the attention weights are not calculated as a dot-product between the query and the keys but are predicted by a linear layer from the query.

Let $z_{HES}^{0}$ be the HES class token embedding (initialized with the shared learnable class token) following the GLN.
Similarly, let $z_{IHC_{i}}^{0}$ be the class token embedding from IHC $i$.
The multi-modal features fusion starts by predicting IHC attention weights with a linear layer from the HES class token embedding:
\begin{equation}w = A z_{HES}^{0} + b\label{equation:w} \end{equation}
where $A$ and $b$ are the weights and bias of the linear layer respectively.
This linear layer outputs a $k$-dimensional vector, with $k$ the number of IHC stains ($k=3$ in our case).
We then perform a weighted sum of the IHC class token embeddings using these weights and add it to the HES one:
\begin{equation} z_{B} =\lambda z_{HES}^{0} + (1 - \lambda)\sum_{i = 1}^{k}w_{i}z_{IHC_{i}}^{0} \end{equation}
where $\lambda$ is a trade-off coefficient between HES and IHC features (which is manually set in our experiments).
The output of multi-modal features fusion is a vector representation of the bag $B$.
This vector is then fed to a classification head for subtype classification.

In the first stage, the multi-modal model is trained to minimize the cross-entropy loss with the ground-truth of the WSI from which the bag is sampled. 

\subsection{Knowledge Distillation from multi to mono modality WSI classification}

We utilize KD to transform the multi-modal model, initially trained with both IHC and HES data (teacher), into a mono-modal one that is solely reliant on HES data (student).
During the training of the student, we feed the teacher the bag of sequences, while the student predicts the subtype from the HES sequence only.
Because the student doesn't make use of IHC we remove the multi-modal features fusion mechanism; indeed, the HES class token ($z_{HES}^{0}$) after the LN is then directly fed to the classification head.
After the teacher model has finished training, it produces outputs that are the corresponding probability predictions for each class.
Subsequently, the student model begins its training stage.
The student model's progress is supported by leveraging the outputs from the teacher model in addition to the ground-truth.
The objective during the training phase of the student model is to minimize the cross-entropy between its own predictions and those of the teacher model, as well as between its predictions and the actual target.
In this study, we implement knowledge distillation in the form of hard distillation, indicating that we treat the teacher model's prediction as a hard target for the student model's training.
If $f^{t}$ and $f^{s}$ denote the teacher and student models respectively, $X^{t}$ and $X^{s}$ their corresponding inputs and $y$ the ground-truth, the student model loss function is defined as:
\begin{equation} Loss = \beta \mathcal{L}(y, f_{s}(X^{s})) + \gamma \mathcal{L}(f_{t}(X^{t}), f_{s}(X^{s})) \end{equation}
where $\mathcal{L}$ is the cross-entropy loss function, $\beta$ and $\gamma$ are coefficients for the ground-truth and teacher supervisions respectively.
We convert the teacher's output to hard labels and apply label smoothing \cite{Mller2019WhenDL} with a parameter of $\alpha=0.1$ to both the teacher's prediction and the true label.

\section{Experiments}
\label{experiments}

\subsection{Data description and processing}

We conduct our experiments on a retrospectively collected cohort of WSIs from patients of the Lille University Hospital diagnosed with DLBCL between 2010 and 2020.
The collection has been meticulously inspected by a pair of pathologists, senior and junior.
Subtyping of the patients was conducted in three distinct ways:
\begin{itemize}
	\item pathologist diagnosis based on visual examination of the HES WSI
	\item application of the Hans algorithm \cite{Hans2004ConfirmationOT}
	\item execution of RT-MLPA \cite{Bobe2017DeterminationOM}
\end{itemize}
However, given that RT-MLPA is recognized as the gold standard for DLBCL subtyping, we take its result as the ground-truth for the subtyping of patients.

Our initial dataset consisted of 204 patients, including 94 diagnosed with the ABC subtype, 66 with GCB, and 44 who did not fall into either category (being either \textit{unclassifiable} or \textit{primary mediastinal large B-cell lymphoma}).
We have excluded these 44 patients from our dataset.
Additionally, 3 patients were removed due to the absence of discernible cancerous tissue in their HES WSI, which is crucial for accurate analysis
For the training phase, we selected 93 patients out of the remaining 157 subjects.
For these patients, we were able to successfully match the HES WSI with their corresponding IHC WSIs (BCL6, CD10, and MUM1).
During the first stage, the teacher is trained with the entire bag of sequences containing all modalities, while in the second stage (KD), the student is trained with the HES sequences only.
For the testing phase, we include the remaining portion of the dataset, which consists of 64 patients with a single HES WSI.
The train set patients are randomly separated into training and validation according to a 80:20 split.

All slides were scanned with a Ventana iScan HT at $\times$40 magnification (0.25$\mu m$ per pixel).
For data pre-processing, the tissue was separated from the background through gaussian blurring and Otsu thresholding segmentation.
For each bag and each modality, we consider sequences of patches with $S=256$ and $p=32$ extracted at $\times$20 magnification.
Our dataset covers 38 different tissue types such as: lymph node (42\%), brain (13\%), mediastinum (5\%), soft tissue (5\%), bone (4\%), etc.
This large diversity of tissue location results in very heterogeneous image characteristics among DLBCL images.

Table \ref{table:data} provides a summary of the distribution of sequences by modality and the distribution of patients in the training and testing splits.

\begin{table}[h]
	\caption{Data summary: number of patients for each set. Training patients were split into 80:20 for train and validation.}
	\centering
	\normalsize
	\begin{tabular}{|l|cc|cc|r|}
		\hline
		& \multicolumn{2}{c|}{train} & \multicolumn{2}{c|}{test} & \multicolumn{1}{r|}{\textbf{total}} \\
		& \multicolumn{1}{c}{ABC} & \multicolumn{1}{c|}{GCB} & \multicolumn{1}{c}{ABC} & \multicolumn{1}{c|}{GCB} & \multicolumn{1}{r|}{} \\
		\hline
		\multicolumn{1}{|l|}{\textbf{sequences}} & \multicolumn{2}{c|}{} & \multicolumn{2}{c|}{} & \multicolumn{1}{r|}{} \\
		\multicolumn{1}{|l|}{\textit{BCL6}} & 83K & 33K & & & \textbf{116K} \\
		\multicolumn{1}{|l|}{\textit{CD10}} & 82K & 33K & & & \textbf{115K} \\
		\multicolumn{1}{|l|}{\textit{MUM1}} & 80K & 35K & & & \textbf{115K} \\
		\multicolumn{1}{|l|}{\textit{HES}} & 78K & 35K & 34K & 32K & \textbf{179K} \\
		\hline
		\hline
		\textbf{patients} & 59 & 34 & 33 & 31 & \textbf{157} \\
		\hline
	\end{tabular}
	\label{table:data}
\end{table}

\subsection{Experimental settings}

For both teacher and student models we use a ViT encoder (d=128, n=3, h=2).
The classification head is a trainable linear layer with two output neurons (one for each subtype), and a softmax activation to transform its output to a probability distribution over the subtypes.
Both the teacher and student models are trained from scratch.
To determine a patient subtype, we apply major voting on all the sequences prediction coming from this patient.
More specifically, we exploit our sequence-based classification model to predict a class for each sequence of the WSI (each sequence having a length of 256 patches), and then the subtype corresponds to the most dominant class over all the predictions.
The teacher is trained with a batch of 32, where each sample of the batch is a bag of sequences ($B$).
We use an initial learning rate of $8.10^{-5}$ with a learning rate decay of 0.5 every 5 epochs.
The student is trained with a batch of 64 and a constant learning rate of $10^{-5}$.
The loss coefficients of the student were set according to best validation accuracy, with $\beta = 0.5$ and $\gamma = 0.5$.
Similarly, the trade-off coefficient of the multi-modal features fusion of the teacher $\lambda$ was set to 0.5.
For both models, we adopt the Adam optimizer with clipping gradient (max norm of 3 for teacher and 5 for students), and apply label smoothing ($\alpha=0.1$) to the ground-truth.
The models are trained for 100 epochs, and the weights with the lowest validation loss are saved.
We apply class balancing by trimming the dominant class, it has shown to improve the validation results.
All models are trained and tested with PyTorch 1.8 on 4 NVIDIA Tesla V100 GPUs, and the metrics are calculated with the scikit-learn (v. 1.1.2) package.
For 100 epochs, the teacher and student training takes 17 and 16 hours respectively.

To analyze and compare the performance of our method we use the precision (PRE), recall (REC) and total accuracy (ACC) which corresponds to the patients correctly classified.

\subsection{Comparison with state-of-the-art methods and models}

\begin{table*}[h]
	\normalsize
	\caption{Comparison of the Student model with state-of-the-art methods for WSI classification on test set (64 patients). We estimated the number of floating operations (number of op.) for a WSI diagnosis based on 5 runs.}
	\centering
	\setlength{\tabcolsep}{7pt}
	 \begin{tabular}{lcccccccc|cc}
		 \hline
		& \multicolumn{2}{c}{ABC} & \multicolumn{2}{c}{GCB} &  \\
		\cmidrule(lr){2-3}\cmidrule(lr){4-5}
		& PRE & REC & PREC & REC & ACC $\uparrow$ & number of op. ($\times 10^{12}$) & params. \\
		 \hline
		HIPT* \cite{Chen2022ScalingVT} & 0.52 & 1.00 & 0.00 & 0.00 & 0.52 & -- & 25M \\
		RSP+CR* \cite{Srinidhi2021SelfsupervisedDC} & 0.54 & 0.94 & 0.71 & 0.16 & 0.56 & 30.8 & 11.8M \\
		MS-DA-MI \cite{Hashimoto2020MultiscaleDM} & 0.58 & 0.76 & 0.62 & 0.42 & 0.59 & 39 & 67M  \\
		KimiaNet* \cite{Riasatian2021FineTuningAT} & 0.58 & 0.85 & 0.69 & 0.35 & 0.61 & 11.5 & 7M \\
		CTransPath* \cite{Wang2022TransformerbasedUC} & 0.71 & 0.73 & 0.70 & 0.68 & 0.70 & 25 & 28M \\
		DTFD-MIL \cite{Zhang2022DTFDMILDF} & 0.68 & 0.85 & 0.78 & 0.58 & 0.72 & 18.6 & 9M \\
		\textbf{ours (student)} & 0.72 & 0.85 & 0.80 & 0.65 & \textbf{0.75} & 0.5 & 0.89M  \\
		\hline
		\multicolumn{6}{p{251pt}}{* indicates pretrained models}  \\
		\multicolumn{8}{p{300pt}}{the abbreviations are: precision (PRE), recall (REC) and accuracy (ACC)}
	\end{tabular}
	\label{tab:stateoftheart}
\end{table*}

We conduct a comparative analysis with state-of-the-art methods in WSI classification.
For this, we have selected several notable works for this purpose including:
\begin{itemize}
	\item HIPT \cite{Chen2022ScalingVT} is composed of hierarchical ViTs with self-supervised pretraining on 10,678 WSI from the TCGA open dataset.
	\item RSP-CR \cite{Srinidhi2021SelfsupervisedDC} pretrained a ResNet-18 with self-supervision on 69 WSIs. The fine-tuning phase involves consistency regularization.
	\item CTransPath \cite{Wang2022TransformerbasedUC} uses a hybrid CNN-Transformer architecture with self-supervised pretraining on 32,220 WSIs from the TCGA and PAIP datasets. The technique employs semantically relevant contrastive learning for unsupervised pretraining.
	\item MS-DA-MIL \cite{Hashimoto2020MultiscaleDM} introduces a multi-scale domain adversarial training approach for WSI classification using a CNN with MIL.
	\item DTFD-MIL \cite{Zhang2022DTFDMILDF} extends the MIL approach for WSI classification by introducing pseudo-bags with attention-based MIL.
	\item KimiaNet \cite{Riasatian2021FineTuningAT} fine-tuned an ImageNet pretrained DenseNet-121 on 7,126 WSIs from the TCGA dataset with 32 cancer classes.
\end{itemize}

For training, we use the pretrained weights provided by the authors for HIPT, RSP-CR and CTransPath, as well as the fine-tuned weights for KimiaNet.
For the later, we reproduce the same downstream approach in \cite{Riasatian2021FineTuningAT} (i.e. we fit a cubic SVM on top of its features).
We apply the specific hyperparameters of all methods from their respective paper.
We have evaluated the six methods on our dataset, ensuring the same data split rules as our method.
Results are presented in Table \ref{tab:stateoftheart}.

According to the metrics we have evaluated, our approach demonstrates superior or comparable results in most cases, when measured against the state-of-the-art methodologies.
While CTransPath displays a marginally better ability to correctly identify GCB patients, with a 3\% higher recall rate, our model consistently delivers a more balanced performance.
Our F-score, a measure that includes both precision (how many selected items are relevant) and recall (how many relevant items are selected), is higher by 2\%.
This implies that our method is better at both correctly identifying positive cases and minimizing the number of false positives, providing a more reliable and accurate overall performance.
Despite their pretraining on large datasets, models like CTransPath, HIPT, or KimiaNet are still outperformed by our method in effectively distinguishing between the two subtypes, with respective accuracy gaps of +5\%, +23\%,and +14\%.
DTFD-MIL, which shows the best results among all comparison methods, is outperformed by our method by +3\%.
Furthermore, it requires on average $25 \times$ more floating operations and contain 10$\times$ more parameters than our student model.
We observe that most methods tend to misclassify GCB samples.
Indeed, among the six compared methods, four ones (namely HIPT, RSP-CR, MS-DA-MIL and KimiaNet) gave for the GCB class a recall value lower than 0.5.
KimiaNet and DTFD-MIL display a comparable ABC recall than our method (0.85 each), but have inferior GCB recall (0.35, 0.58 and 0.65 respectively).
Additionally, we observe that our model is very light compared to other methods, with less than one million parameters, whereas other models contain between 7 and 67 million parameters.

These models face challenges in accurately distinguishing between DLBCL subtypes due to the specific nature of this task.
Indeed, because the task is highly complex due to the presence of visually similar features among the subtypes, these methods also do not escape this complexity and struggle to effectively capture or adapt to it during their training process.
On the other hand, by leveraging different data modalities, our teacher model is able to learn a richer set of features better suited to the task at hand.
The teacher model then provides a crucial supervision to guide the student model to discover the most relevant features for the given task.

\begin{figure*}[h]
	\centering
	\begin{subfigure}[t]{0.46\textwidth}
		\includegraphics[trim={0 0 0 1cm},width=\linewidth]{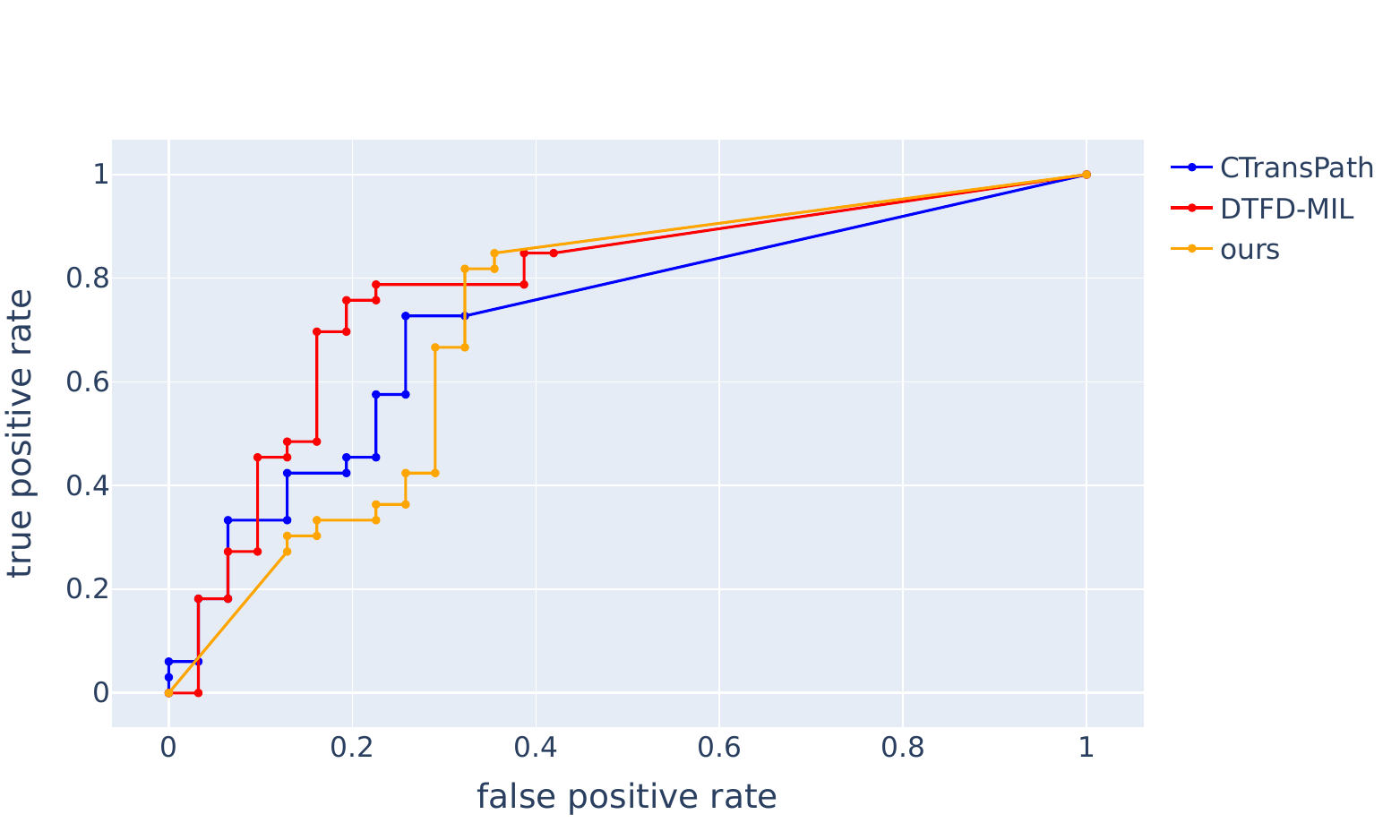}
		\caption{ROC curves of the ABC class.}
		\label{fig:rocABC}
	\end{subfigure}
	\begin{subfigure}[t]{0.46\textwidth}
		\includegraphics[trim={0 0 0 1cm},width=\linewidth]{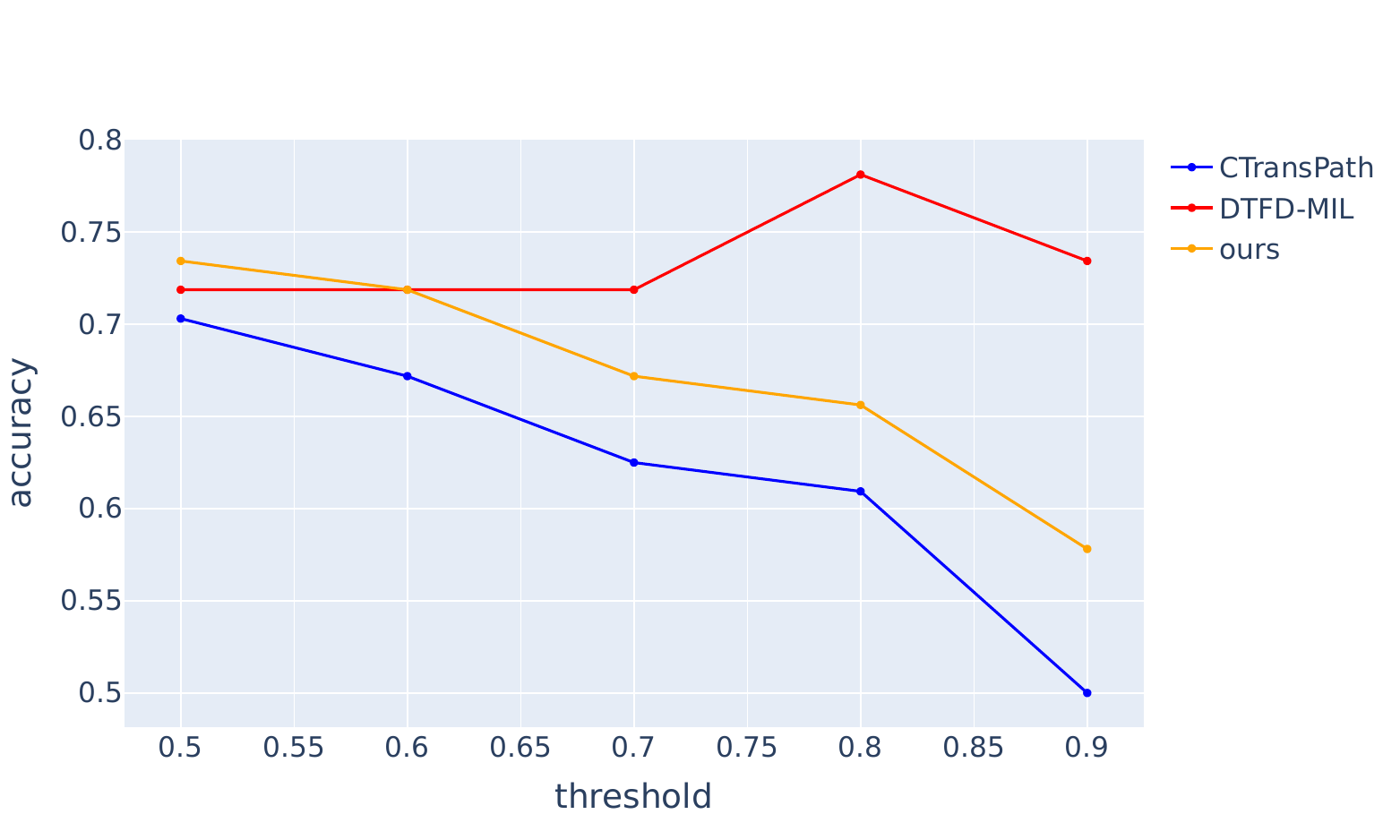}
		\caption{Accuracy with respect to classifier threshold of the ABC class.}
		\label{fig:accABC}
    \end{subfigure}
	\begin{subfigure}[t]{0.46\textwidth}
		\includegraphics[width=\linewidth]{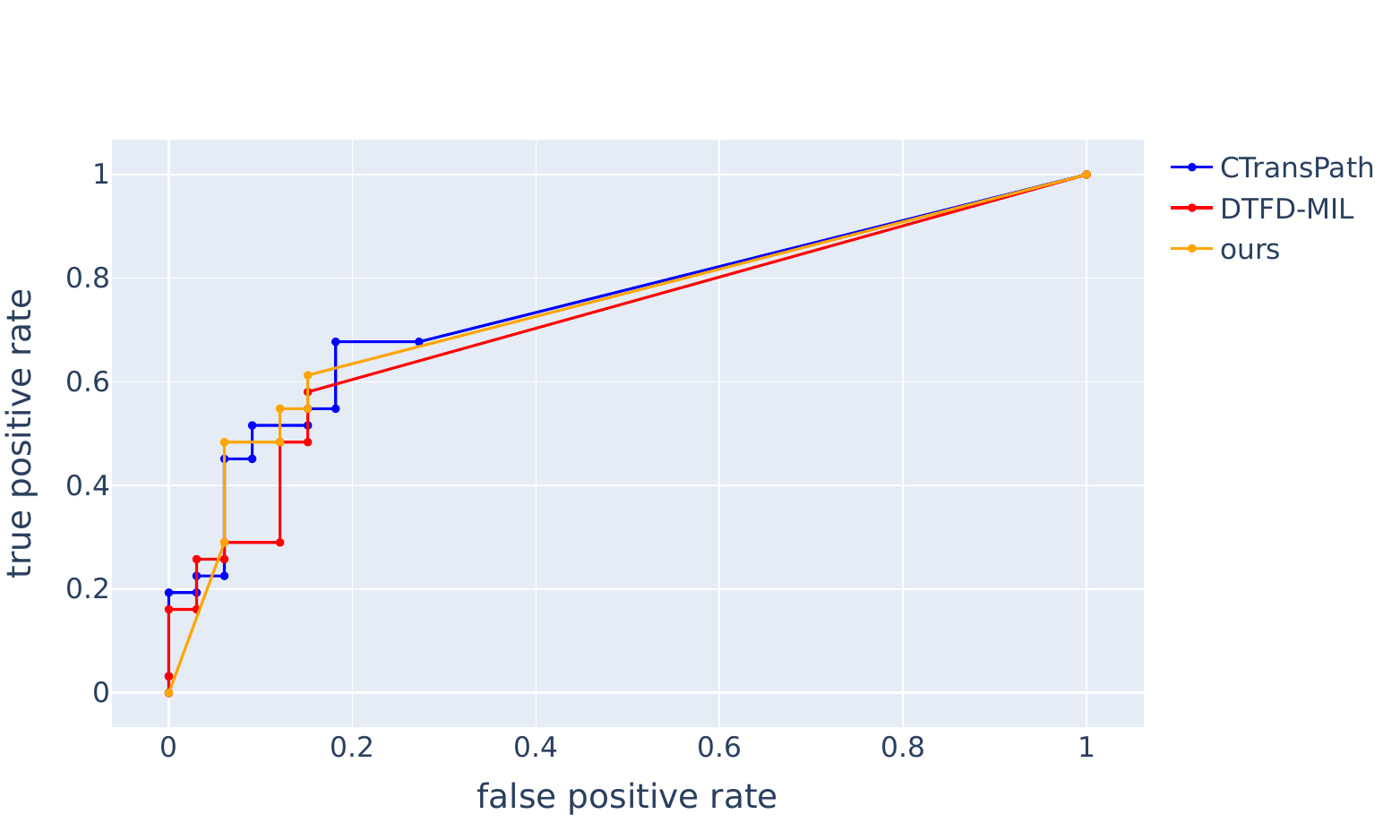}
		\caption{ROC curves of the GCB class.}
		\label{fig:rocGCB}
    \end{subfigure}
	\begin{subfigure}[t]{0.46\textwidth}
		\includegraphics[width=\linewidth]{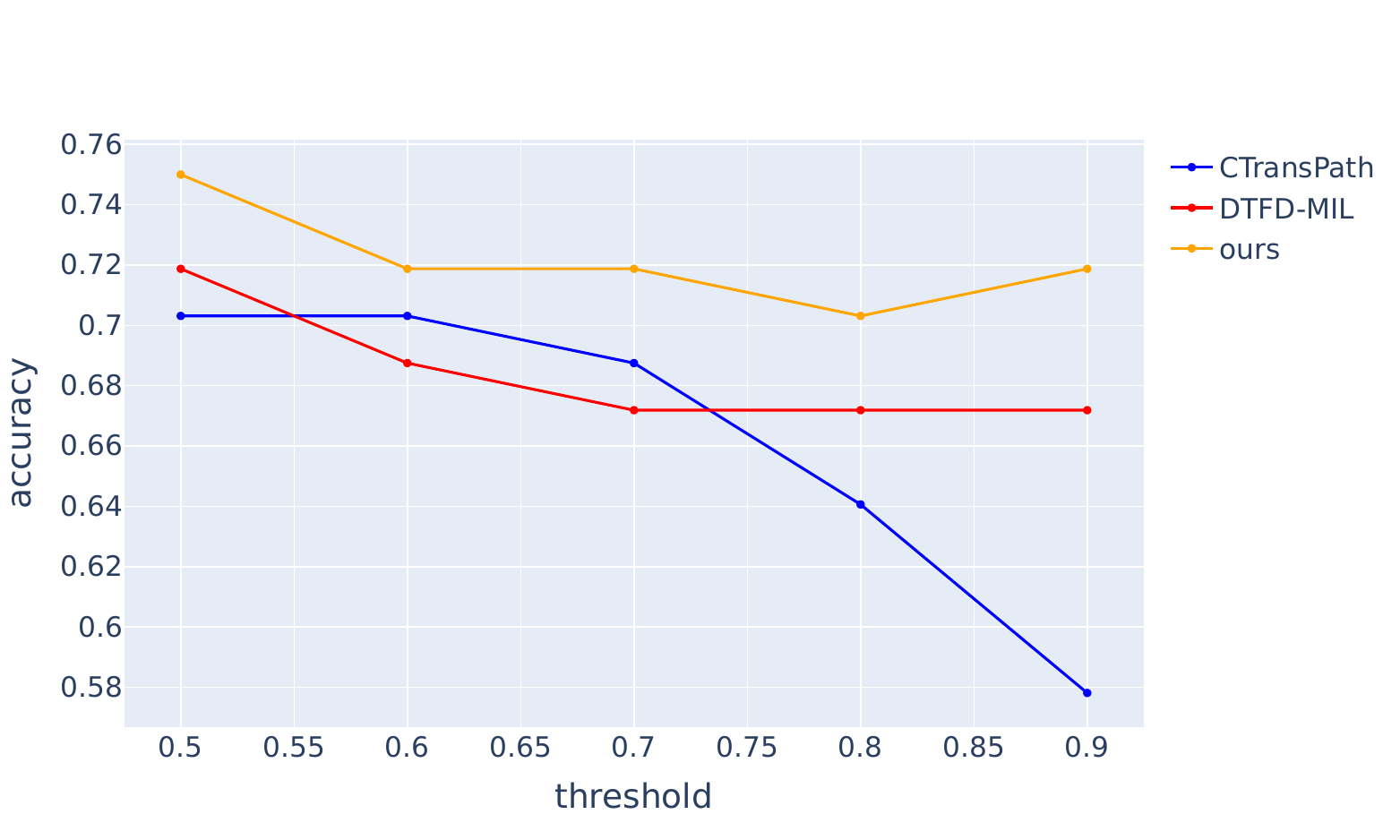}
		\caption{Accuracy with respect to classifier threshold of the GCB class.}
		\label{fig:accGCB}
    \end{subfigure}
	\caption{Performance comparisons of state-of-the-art methods obtained on the test set.}
\end{figure*}

Furthermore, we generate ROC curves, for each class separately, for our student model and the two methods in Table \ref{tab:stateoftheart} with an accuracy higher than 70\%.
The curves for the ABC class are shown in Figure \ref{fig:rocABC}.
We observe that DFTD-MIL shows relatively higher true positive rates (TPR) for low false positive rates (FPR); see FPR in the range [0, 0.3] in the Figure \ref{fig:rocABC}.
However, our method shows the best TPR when the FPR exceeds 0.3.
To more explicitly illustrates the sensitivity of the methods with respect to the classification threshold, we have calculated the accuracy of the classifiers for this class, across several thresholds (in the range [0.5, 1]), displayed in Figure \ref{fig:accABC}.
We observe that at a threshold of 0.5, our method shows the best accuracy (0.75), however, contrary to DTFD-MIL, its accuracy decreases with higher thresholds.
Overall, our method still exhibits for this class the best F1-score (0.78) compared to DTFD-MIL (0.76) and CTransPath (0.72).
Similarly, the ROC curves for the GCB class are shown in Figure \ref{fig:rocGCB}.
We observe that the three methods are quite competitive.
Nevertheless, as shown in Figure \ref{fig:accGCB}, our method displays a better sensitivity in term of classification threshold, since its accuracy outperforms those of the other methods whatever the threshold considered.
Moreover, similarly to the ABC class, the F1-score of our method (0.71) is greater than DTFD-MIL (0.67) and CTransPath (0.69) for the GCB class.

\subsection{Comparison with medical diagnosis techniques}

In Table \ref{tab:diagnosistechniques}, we present a comparative analysis with diagnosis techniques.
Here, the student model, trained with our method, is compared with the visual analyses of two pathologists and the Hans algorithm (presented in section \ref{sec:introduction}).
On HES images, our model outperforms the two pathologists in distinguishing between the two subtypes, achieving an accuracy that is higher by a margin of at least 9\%.
The relatively low diagnosis accuracy reached by the pathologists is explained by the fact that they performed their analysis on a single image modality namely HES without any prior knowledge on the IHC modalities.
In this sense, we observe that when the pathologists exploit the three IHC modalities, via the Hans algorithm, they reach a better diagnosis accuracy of 83\%.
Nevertheless, as presented in Table \ref{tab:fnr}, our model demonstrates a lower misclassification rate for ABC patients compared to all other diagnostic techniques."
This means that a patient with ABC subtype (subtype with the lowest survival chances) is less likely to be misclassified by our model than the Hans algorithm.

\begin{table*}[h]
	\centering
	\normalsize
	\caption{Comparison of the Student model with clinical diagnosis techniques for DLBCL subtyping on test set (64 patients).}
	\setlength{\tabcolsep}{5pt}
	 \begin{tabular}{lccccccc}
		 \hline
		& \multicolumn{3}{c}{ABC}& \multicolumn{3}{c}{GCB}&  \\
		\cmidrule(lr){2-4}\cmidrule(lr){5-7}
		& PRE& REC& F1& PREC& REC& F1& ACC \\
		 \hline
		pathologist A& 0.69& 0.67& 0.68& 0.65& 0.65& 0.65& 0.66 \\
		pathologist B& 0.57& 0.64& 0.60& 0.54& 0.45& 0.49& 0.55 \\
		ours (student)& 0.72& 0.85& 0.78& 0.80& 0.65& 0.71& 0.75 \\
		 \hline
		IHC (Hans et al. \cite{Hans2004ConfirmationOT})& 0.92& 0.73& 0.81& 0.76& 0.94& 0.84& 0.83 \\
		\hline
	\end{tabular}
	\label{tab:diagnosistechniques}
\end{table*}

\begin{table}[b]
	\centering
	\normalsize
	\caption{False negative rates.}
	\setlength{\tabcolsep}{3pt}
	\begin{tabular}{lccc}
		\hline
		& ABC& GCB \\
		\hline
		pathologist A& 0.33& 0.35 \\
		pathologist B& 0.36& 0.55 \\
		ours (student)& 0.15& 0.35 \\
		\hline
		IHC (Hans et al. \cite{Hans2004ConfirmationOT})& 0.27& 0.06 \\
		\hline
	\end{tabular}
	\label{tab:fnr}
\end{table}

\begin{figure}[h]
	\includegraphics[trim={2.1cm 0cm 1.8cm 0.5cm}, clip, width=\linewidth]{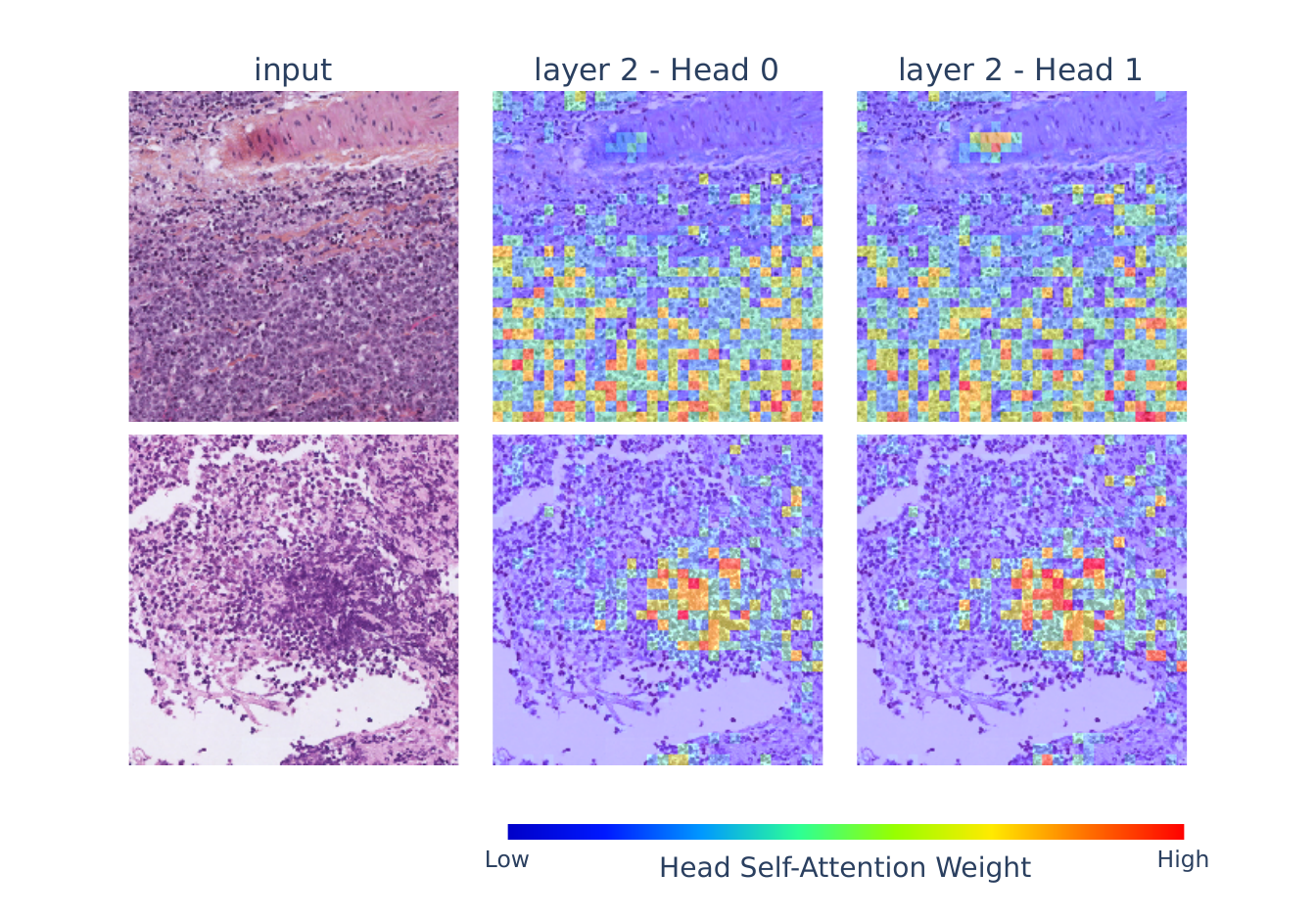}
	\caption{Attention maps generated by our student model on partially tumoral regions.}
	\label{fig:attention}
\end{figure}

We extended our analysis further by visualizing the attention scores of the ViT encoder in the 
last layer (just before the classification head).
Fig. \ref{fig:attention} shows results of these scores obtained on two WSI patients.
One can observe that areas rich in tumoral cells are receiving high attention.

\subsection{Power-law scaling}
\label{section:powerlaw}

\begin{figure*}[h]
	\centering
	\includegraphics[trim={1cm 10cm 1cm 12cm}, clip, width=0.8\linewidth]{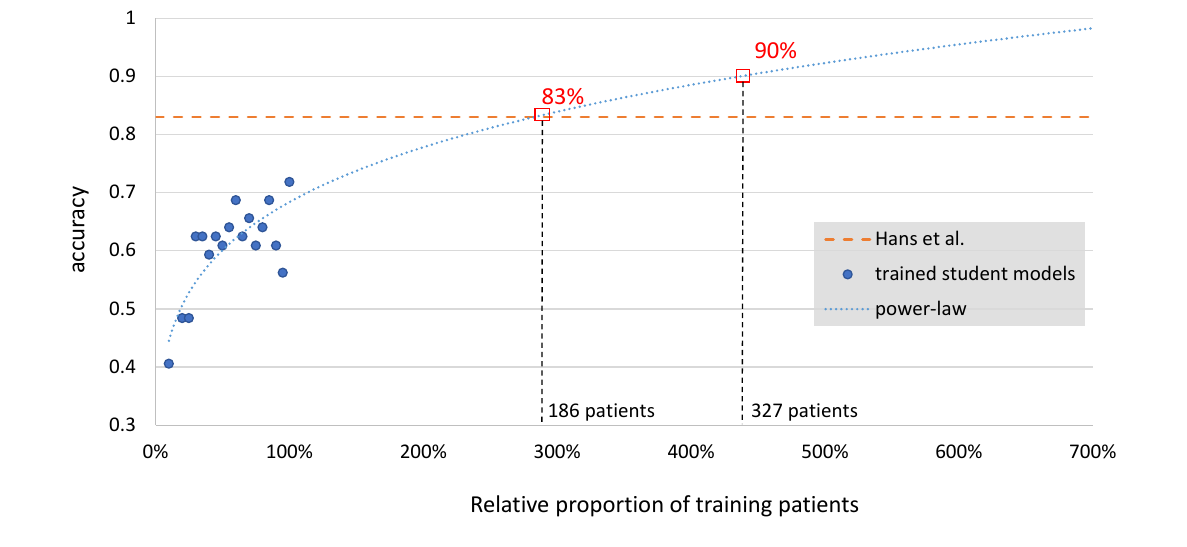}
	\caption{We fit a power-law curve to project the generalization performance of the student model using a larger training set. Accuracy corresponds to the performance of the classification model obtained on the test set of 64 patients.}
	\label{fig:powerlaw}
\end{figure*}

Scaling law in deep learning refers to the model generalization capability when its computational power and/or training data is scaled \cite{Abnar2021ExploringTL}.
It indicates that increasing resources such as the computing power or volume of training data, can considerably improve the model's generalization.
Studies on ViT \cite{Zhai2021ScalingVT} have shown that when we increase both the model size and the amount of training data, the generalization error decreases inevitably.
In our study, we are interested in observing the student generalization performance when training data is scaled up.
In order to achieve this, we trained both teacher and student models on incremental portions of the original dataset, varying from 10\% to 100\%, measured by number of patients.
These subsets are cumulative, meaning each new subset contains all patients from the previous one, with an added 5\% patients of the original training set.
Thus, with every step, the subset grows by retaining the existing patients and incorporating new ones.
Across all subsets, we train the models with the same initial parameters as well as the same hyperparameters, ensuring equal conditions for every trained models.
Observing Fig. \ref{fig:powerlaw}, the model accuracy exhibits a general increase with the enlargement of the training data.
However, distinct declines in accuracy are observed at 90\% and 95\%, which could be attributed to the incorporation of patient with images markedly different from the existing dataset.
Despite this, the introduction of more patients afterwards leads to a rebound in accuracy, highlighting the positive impact on the model's performance.
For instance, the power-law curve shows that the student model could outperform the Hans algorithm and reach an accuracy of 90\% with a training set of 327 patients. 

\subsection{Ablation study}

This section outlines a series of experiments conducted to evaluate the significance of various features of our architectural design, with results detailed in Table \ref{tab:ablation}.
Our exploration initiated with a thorough analysis of KD, a critical component for its role in guiding the student model using insights from the teacher model.
We set up an experiment to train the student model exclusively with the true target, without teacher guidance, to evaluate the impact of such self-reliant learning on generalization performance.
To assess the effects of independent learning on the student model generalization abilities, we removed the teacher supervision, therefore solely relying the true target one.
Our investigation uncovered interesting findings: without teacher guidance the overall accuracy decreased by 5\%.
This underscores the crucial influence of KD in enhancing the student model’s performance, highlighting the importance of the teacher-student dynamic in the learning process.
These insights motivate further research into optimizing KD within our architecture to potentially improve the student model's performance further.

To minimize the impact of initial parameters choices on the results, we maintained consistent initial model parameters throughout the following experiments.
By using the same initial parameters, we could fairly assess the effects of the specific features investigated.

In another experimental setup (\textit{without learnable att.}), substituting learnable attention with dot-product attention led to a notable 17\% decrease in accuracy.
This resulted in an overall accuracy score even lower than that achieved without any KD at all (58\% vs. 70\%, respectively).
This discrepancy is likely to be attributed to the variations in color distribution between IHC and HES images.
These variations may impede the dot-product similarity from effectively capturing relevant information.
Therefore, our findings emphasize the crucial role of learnable attention in our methodology.
It allows the model to learn to focus on the most relevant IHC information, which is essential for improving the performance of the model.

Next, we removed the softmax activation of the encoder MLP to evaluate its effect on the model's generalization accuracy (\textit{without softmax}).
The modified model demonstrated a substantial decrease in accuracy by 17\%, underscoring the essential role of the softmax activation in improving classification performance.

We also investigated how class balancing affects model accuracy (\textit{without class balance}). 
We trained both the teacher and student models without class balancing.
The student model's accuracy showed a significant decrease of 13\%.
This suggests that class balancing plays a vital role in the model learning process by ensuring that all classes are equally represented during training.

\begin{table}[h]
	\centering
	\normalsize
	\caption{Ablation results on test set (64 patients). We show the F1-score of each class and the total accuracy.}
	\setlength{\tabcolsep}{6pt}
	 \begin{tabular}{lccc}
		\hline
		& \multicolumn{2}{c}{F1} & ACC  \\
		\cmidrule(lr){2-3}
		& ABC & GCB & \\
		\hline
		Original (student) model & \textbf{0.78} & \textbf{0.71} & \textbf{0.75} \\
		\hline
		without softmax & 0.54 & 0.61 & 0.58 \\
		without class balance & 0.62 & 0.57 & 0.59 \\
		without learnable att. & 0.61 & 0.64 & 0.63 \\
		without KD & 0.73 & 0.67 & 0.70 \\
		\hline
	\end{tabular}
	\label{tab:ablation}
\end{table}

\subsection{Results on external data}

To show the generalization capability of our method, we conduct an experimental study on an external dataset.
The dataset is named BCI \footnote{available at \url{https://bupt-ai-cz.github.io/BCI/}} (Breast Cancer Immunohistochemical) \cite{Liu2022BCIBC} and is composed of two modalities (HES and IHC) with 4,873 images each.
It is dedicated to the the evaluation of human epidermal growth factor receptor 2 (HER2) expression, an essential factor to formulate a precise treatment for breast cancer \cite{Liu2022BCIBC}.
The dataset offers the ground-truth about the expression level of the HER2 factor for each pair of images.
In this frame, two recent methods \cite{Shovon2022StrategiesFE,Wang2023HAHNetAC} of the state-of-the-art have been proposed to address the classification task of the HES images among four different levels of expression.
The dataset meets our needs in term of multiple modalities, which can be exploited by our approach to transfer the knowledge from a HES+IHC classifier to a HES one.
However, the images provided in the dataset are not high-resolution WSI as in our case, but rather images of 1024x1024 pixels representing a tiny region of the patient tissue.
Therefore, the Vision Transformer-based backbone of our architecture, is not suited for analyzing such form of data, in the sense that the backbone requires in input a sequence of patches uniformly sampled from a large region of tissue of the WSI.
Hence, to adapt our architecture to the BCI dataset, we have set up a variant in which we replaced the Transformer-based backbone by a CNN-based one while preserving the core components of our approach, which are the Knowledge Distillation and the multi-modal features fusion mechanisms.

For the choice of the CNN backbone, we exploited the two most competitive methods on our original dataset, namely DTFD-MIL and CTransPath (see Table \ref{tab:stateoftheart}), which permitted to produce three models.
To analyze the performance of these models, we also considered in our experimental study the remaining methods of the Table \ref{tab:stateoftheart}; except HIPT and MS-DA-MI, for which their input requirements are not compliant with the BCI dataset.
All the methods have been trained accordingly to the settings reported in their original papers.
In addition, to highlight the competitiveness of our approach, we also included the results of the state-of-the-art methods (HAHNet \cite{Wang2023HAHNetAC} and HE-HER2Net \cite{Shovon2022StrategiesFE}) that have been specifically developed for the BCI dataset.

Table \ref{tab:bci} summarizes the results obtained on the same test set.
The table shows that our three variants improve the accuracy of the original methods with a margin ranging from 3\% to 8\%.
Moreover, we can observe that our variant (based on CTransPath) and the HAHNet model from the state-of-the-art obtained the best accuracy (0.93).

\begin{table}[h]
	\centering
	\normalsize
	\caption{Comparison of original state-of-the-art methods and our variant on the BCI \cite{Liu2022BCIBC} test set.}
	\setlength{\tabcolsep}{3pt}
	\begin{tabular}{llc}
		\hline
		& & Accuracy \\
		\hline
		DTFD-MIL (VGG-19) \cite{Zhang2022DTFDMILDF}& original& 0.71 \\
		& \textbf{our variant}& 0.74 \\
		\hline
		DTFD-MIL (ResNet-50) \cite{Zhang2022DTFDMILDF}& original& 0.83 \\
		& \textbf{our variant}& 0.86 \\
		\hline
		CTransPath \cite{Wang2022TransformerbasedUC}& original& 0.85 \\
		& \textbf{our variant}& \textbf{0.93} \\
		\hline
		RSP+CR \cite{Srinidhi2021SelfsupervisedDC}& --& 0.79 \\
		\hline
		KimiaNet \cite{Riasatian2021FineTuningAT}& --& 0.82 \\
		\hline
		HE-HER2Net \cite{Shovon2022StrategiesFE}& --& 0.87 \\
		\hline
		HAHNet \cite{Wang2023HAHNetAC}& --& \textbf{0.93} \\
		\hline
	\end{tabular}
	\label{tab:bci}
\end{table}

\section{Conclusion}
We proposed a new methodology based on knowledge distillation to distill a multi-modal vision transformer-based classifier into a mono-modal one for DLBCL molecular subtyping on WSI.
We demonstrated the effectiveness of our methodology through an experimental study on 157 patients.
Our mono-modal model outperformed state-of-the-art methods, demonstrating the benefit of leveraging multi-modal model knowledge during training.
Furthermore, this benefit has been confirmed through an additional experimental study on a external breast cancer dataset (BCI dataset \cite{Liu2022BCIBC}).
Inspired by power-law scaling in Vision Transformers, we also explored how our mono-modal model's diagnostic accuracy scales within larger training data scenarios.
Our findings indicate that, with a reasonable number of additional patients, our model can compete with standard clinical methods like IHC \cite{Hans2004ConfirmationOT}.
Because of the difficulty of DLBCL subtyping on HES slides, for the same patient, pathologists often rely on several slides to make a diagnosis.
Thus, our methodology could be extended to incorporate these HES slides in the student model training to increase the training data.
Finally, active learning-driven data selection has been shown to enhance the generalization of deep learning models while minimizing the volume of data required.
It could be leveraged to acquire more data efficiently with the objective of reaching competitive performance with the Hans algorithm.

\section*{Data Protection}
The material for conducting the experiments of this work was declared to the Data Protection Officer (DPO) of CHU de Lille and the National Commission on Informatics and Liberty (CNIL) (declaration number 1129).
Furthermore, all the patients were anonimized.

\bibliographystyle{ieeetr}
\bibliography{references}

\end{document}